\documentclass[12pt]{article}

\usepackage{times}
\usepackage{amsmath}
\usepackage{graphicx}
\usepackage{upgreek}

\title{A mirrorless spinwave resonator}
\author{Olivier Pinel$^+$, Jesse L. Everett$^+$, Mahdi Hosseini, Geoff T. Campbell, \and Ben C. Buchler* and Ping Koy Lam* \\
\normalsize{Centre for Quantum Computation and Communication Technology,}\\ \normalsize{The Australian National University, Canberra ACT 0200, Australia.} \\\normalsize{*Corresponding authors: Ben.Buchler@anu.edu.au (BCB), Ping.Lam@anu.edu.au (PKL).}\\\normalsize{+ these authors contributed equally to this work}}
\begin{document}
\maketitle
\begin{abstract}
Optical resonance is central to a wide range of optical devices and techniques.  In an optical cavity, the round-trip length and mirror reflectivity can be chosen to optimize the circulating optical power, linewidth, and free-spectral range (FSR) for a given application.  In this paper we show how an atomic spinwave system, with no physical mirrors, can behave in a manner that is analogous to an optical cavity.  We demonstrate this similarity by characterising the build-up and decay of the resonance in the time domain, and measuring the effective optical linewidth and FSR in the frequency domain.  Our spinwave is generated in a 20 cm long Rb gas cell, yet it facilitates an effective FSR of 83 kHz, which would require a round-trip path of 3.6 km in a free-space optical cavity.  Furthermore, the spinwave coupling is controllable enabling dynamic tuning of the effective cavity parameters.
\end{abstract}

Optical resonators are an essential technology for photonic systems. For example, they are used for stabilizing optical frequencies \cite{Schiller2004,Ludlow2007,Millo2009,Kessler2012,Cole2013}, filtering optical fields \cite{TheVirgocollaboration2007,Hockel2010,Palittapongarnpim2012}, enabling non-linear optics \cite{boyd2003nonlinear} and precision measurement \cite{berden2009cavity,Abbott2009,Jones2005}.  There has always been significant interest in developing novel and, in particular, compact optical resonators \cite{Vahala2003} due to the innumerable applications they enable.  In the last decade, for example, resonator design has been extended to include mapping optical field into metallic plasmon resonances \cite{Ditlbacher2005,Min2009,Bozhevolnyi2006,Rottler2013}. The wide range of physical shapes and materials used in the construction of the resonators leads to an equally wide range of optical characteristics, although narrow resonant linewidth remains a common goal regardless of the design.  Here, we present a new type of resonator in which optical coherence is stored in an ensemble of atoms.  The combined mode of the light and atomic coherence that carries the information during storage is best described as a spinwave-polariton \cite{Fleischhauer2002,Fleischhauer2005,Hetet2008d}. The behavior of the stored spinwave is highly tunable, allowing real-time control of parameters such as linewidth and free-spectral range (FSR).

The key component of a standard optical resonator is a partially reflecting mirror, or beamsplitter, that couples light into and out of the cavity.  In a resonant optical cavity, light incident on the input mirror interferes with the light circulating in the cavity to constructively enhance the cavity mode.  Previous modeling \cite{Longdell2008} and experimental work \cite{Campbell2012b,Reim2012} has shown how coherent atomic memories can be understood as an atomic beamsplitter that acts to transfer excitation between optical and atomic modes.  To build the analogue of a cavity using an atomic beamsplitter, we periodically store pulses of light that are timed to arrive at the memory just as the stored spinwave is being recalled back into an optical field.  This allows the memory storage to act as the mirror to a cavity where the round-trip time in the cavity is now the time between the pulses.  This is the same idea used in resonant cavity ring-down spectroscopy \cite{Lehmann1996,Crosson1999}.  This concept is illustrated in figs. 1C and D for the case where a pulse arrives at the memory on alternate memory recall events.  As we will illustrate, in our experiment the equivalent cavity has a round trip length of 3.6 km.
\section*{Theory: using an atom-light beamsplitter to build a cavity.}
As described in refs. \cite{Longdell2008,Campbell2012b,Reim2012}, the storage and recall of light in an atomic spinwave can be described as beamsplitter operation.  For an optical depth, , the fraction of light transmitted through the memory is given by $T=e^{-2\pi\beta}$ and the fraction stored in the memory, analogous to the fraction reflected from a physical beamsplitter, is given by $R=1-T$.  Suppose we start with a single mode of light with amplitude $A_0$.  When injected into the memory this will be split into a transmitted mode, with amplitude $A_1=\sqrt{T}A_0$, and a spinwave mode, with amplitude $S_1=\sqrt{R}A_0$, where $S$ is the appropriately normalised collective atomic excitation.  By symmetry, if the storage efficiency is given by $R$, then the recall efficiency is also equal to $R$, assuming constant optical depth.  Similarly, on recall the fraction of spinwave that remains in the atomic medium is given by $T$.

Supposing now that we recall the spinwave at time $\tau$, while at the same time injecting a second optical mode with amplitude $A_0$.  The spinwave, $S_2$, after this operation will be constructed by interfering the spinwave arising from the second optical mode and the fraction of the original spinwave that remains after recall.  Using the relations above we may write $S_2=\sqrt{T}S_1e^{-\gamma\tau}e^{i\phi}+\sqrt{R}A_0$ where we have included decay of the spinwave mode at a rate $\gamma$ and a relative phase $\phi$. In a mirror based cavity the phase  depends on the round-trip time of the cavity.  In our experiment this phase depends on the storage time of our memory, which may be chosen arbitrarily. We may continue to recursively define the amplitude of the spinwave mode, $S_n$, for $n$ pulses to arrive at an infinite sum as $n\rightarrow\infty$.  This gives the steady-state value of spinwave amplitude to be:
\begin{equation}
S_\infty=\frac{\sqrt{R}A_0}{1-\sqrt{T}e^{-\gamma\tau}e^{i\phi}}\label{equation1}
\end{equation}

This is a standard result for the steady-state amplitude of the field inside an optical resonator \cite{siegman1986lasers}. The interference behaviour that leads to this result is the same regardless of whether the input light is continuous or pulsed \cite{Lehmann1996,Crosson1999}, provided the pulse bandwidth is less than the resonator bandwidth. The square of Eq. \ref{equation1} gives the familiar Airy-function cavity transmission spectrum as a function of the round-trip phase, $\phi$. In this derivation we have not included any details of the memory bandwidth, which as we will show, spectrally limits the range of the resonant peaks.

The amplitude of a mode transmitted through the memory is obtained by combining the light recalled from the spinwave with the light transmitted through the memory to give $A_t=\sqrt{T}A_0-\sqrt{R}S_n$, noting that the subtraction required for conservation of energy. The spinwave resonator will be impedance matched if this amplitude is equal to zero.  If, after running the experiment for n pulses we wish to measure the amplitude of the spinwave in the memory, we recall from the memory without injecting a pulse.  In this case the recalled optical mode amplitude, $A_t$, will be proportional to the amplitude of the spinwave, $S_n$.
\section*{Experimental Setup and Method}
\begin{figure}
\includegraphics[width=\textwidth]{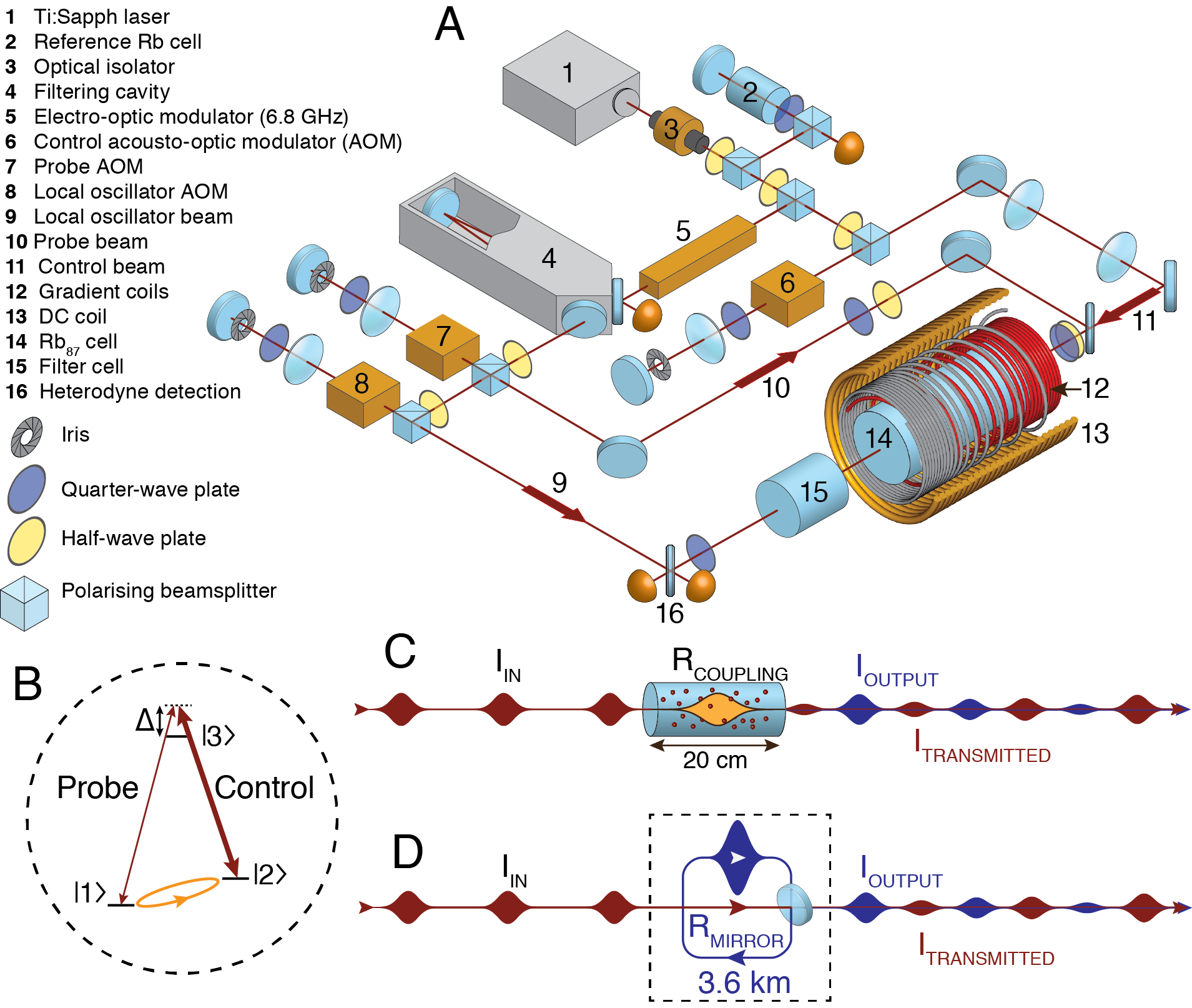}
\caption{\textbf{(A)} Experimental setup showing the laser system. The gradient coils are used to apply opposite sign magnetic field gradients while the DC coils tunes the center frequency of the atomic absorption. \textbf{(B)} The effective three-level atom addressed by the laser system. The detuning $\Delta$ is set to around 2 GHz. Comparison of the \textbf{(C)} spinwave resonator to a \textbf{(D)} ring cavity. The dimensional equivalence of our 20 cm resonator is a ring cavity with a single coupling mirror and a round trip optical path of up to 3.6 km. Our experimental scheme is analogous to sending one pulse into such a ring cavity for every two round trips the built-up pulse makes inside the cavity. The red pulses ($I_{IN}$) pump fresh light into the memory (cavity) and if the phase is correct will constructively interfere into the memory (cavity). If the impedance matching is imperfect, some of this light leaks out ($I_{TRANSMITTED}$). The output pulses ($I_{OUTPUT}$, blue) are withdrawn every second cycle, in the absence of an input pulse. They probe of how much light is stored in the memory (cavity). For a cavity the mirror reflectivity ($R_{MIRROR}$) is fixed. The effective mirror reflectivity in the memory ($R_{COUPLING}$)
may be dynamically controlled.}\label{figure1}
\end{figure}

Our experiment uses a three-level Gradient Echo Memory (GEM) scheme to control the interaction between light and a vapor of Rubidium atoms contained in a warm gas cell.  This memory protocol has been investigated extensively in previous work \cite{Moiseev2008,Alexander2006,Hosseini2009,Hosseini2011,Hosseini2011a,Hosseini2012,Hetet2008} and has been shown to act as a noise-free, high-efficiency quantum memory.  In brief, this system works using a two-photon Raman interaction (see fig 1B.) involving two hyperfine ground-states ($|1>$ , $|2>$) and an excited state $|3>$.  Under the influence of a bright control field, detuned by 2 GHz from the $|2> \rightarrow |3>$ transition, a weak probe pulse, detuned similarly from the $|1> \rightarrow |3>$ transition, is absorbed by the atoms.  The light pulse is then stored as a coherence of the hyperfine ground states.   The bandwidth of the memory is tuned by applying a longitudinal frequency gradient to the atomic ensemble.  This is achieved using gradient coils (see Fig. \ref{figure1}A) that induce a linearly varying Zeeman shift along the length of the atomic cell so that the frequency spectrum of the pulse is mapped onto the spatial axis of atomic coherence.  Memory recall is achieved by reversing the sign of the gradient, which causes rephasing of the atomic coherences and the generation of a photon echo in the forward direction. The strength of the control field may be used to tune the optical depth experienced by the probe passing through the atomic sample. This tunability is crucial in our experiment as it allows us to program the fraction of light that is coupled into the memory.  

The mirror-based analogue to our spinwave system is a ring-cavity \cite{Higginbottom2012}, as shown in Fig. \ref{figure1}D.  By arranging the coupling mirror to reflect into the ring cavity we have correspondence with the notation in Eq. \ref{equation1}, i.e. coupling into the spinwave is equivalent to reflection off the coupling mirror.  In Fig. \ref{figure1}D increasing the transmission of the coupling beamsplitter will increase the finesse of the resonator.  Similarly, increasing the transmission of our memory will increase the finesse of our analogue spinwave resonator.  This is a counterintuitive result in the sense that quantum memory experiments generally struggle to increase optical depth and absorption into the memory.  Here we must deliberately reduce the optical depth to improve performance.

\section*{Results}
\begin{figure}
\includegraphics[width=15cm]{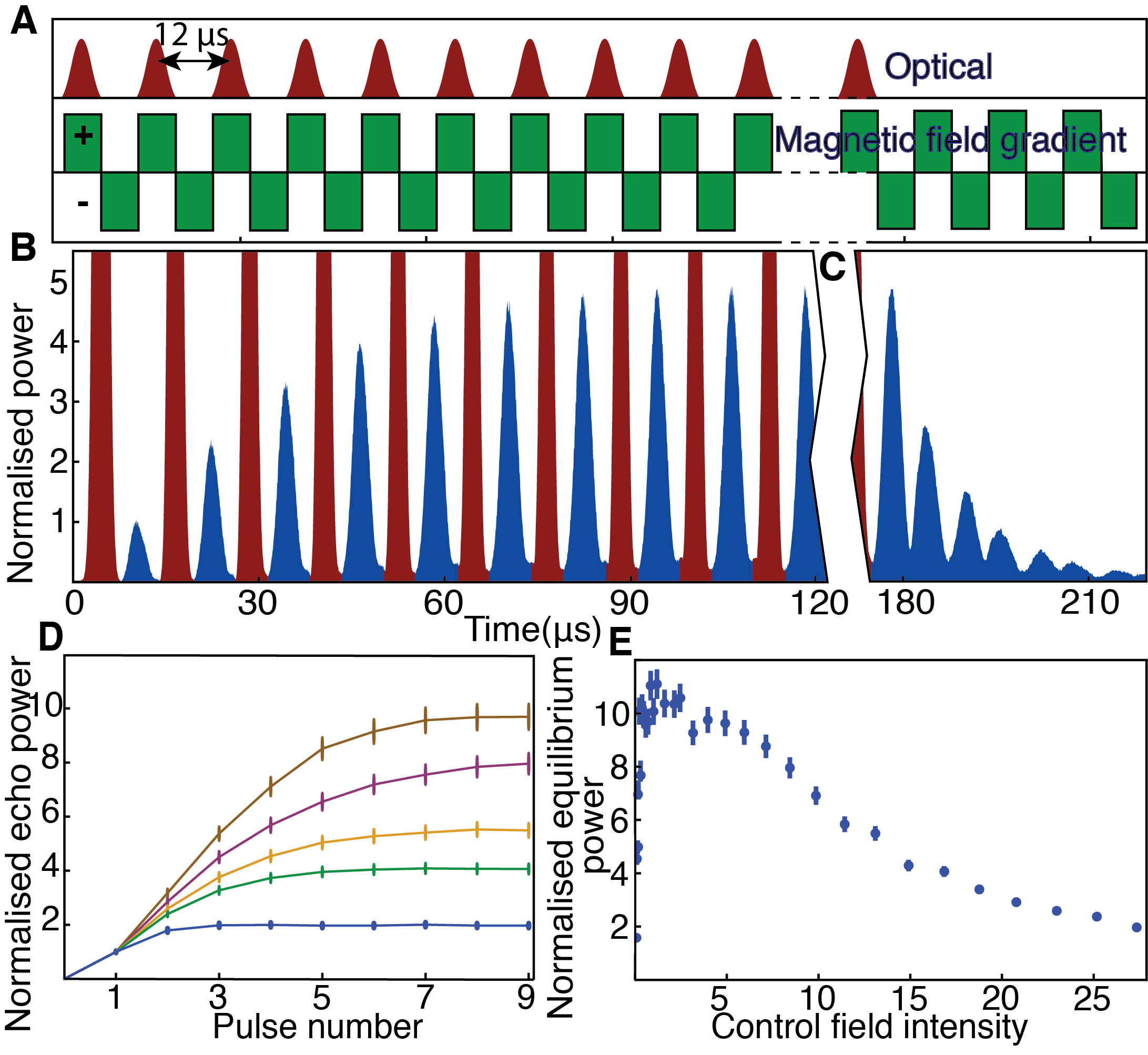}
\caption{\textbf{Spinwave resonator time domain data. (A)} The timing scheme of the input pulses (top) and gradient switching (bottom). A pulse is incident once every full cycle of the gradient, which corresponds to every second recall period. \textbf{(B)} In our experimental data, most of the energy in each input pulse (red) is transmitted through the memory, which saturates the detector. Every second recall period, when no input pulse is incident, we observe successive increases in the recalled light (blue) indicating a growing spinwave in the memory cell. After about 8 recall pulses, equilibrium is established and the recalled pulses remain at a constant energy. \textbf{(C)} With the input pulses switched off, we observe the ring-down of the spinwave resonator. \textbf{(D)} The echo power, scaled to the initial echo, demonstrating the accumulation of the spinwave for control field irradiance, in descending order of final echo power, 0.7 mW.cm$^{-2}$, 8.4 mW.cm$^{-2}$, 13.1 mW.cm$^{-2}$, 16.8 mW.cm$^{-2}$, and 25.2 mW.cm$^{-2}$. The lines are a guide only. \textbf{(E)} Equilibrium echo intensity relative to initial echo as a function
of control field intensity.}\label{figure2}
\end{figure}

To show constructive interference into the spinwave, we periodically switch the magnetic field gradient while injecting new pulses into the memory at times when the spinwave is rephased.  As shown in Figs. \ref{figure2}A and B, pulses are sent into the memory on every second recall period. In between the injection of fresh pulses, the echoes retrieved show accumulation of energy until equilibrium is reached, where losses balance the incident energy. 

The ‘ring-down’ of a cavity describes what is observed when a resonator is filled with light that is then allowed to leak out.  Cavity ring-down measurement has applications to spectroscopy and is an established technique for sensitive absorption measurements \cite{berden2009cavity,Lehmann1996,Crosson1999}.  In the absence of the input light, we observe ring-down behaviour in our system, as shown in Fig. \ref{figure2}C. This is analogous to a pulse travelling around a ring cavity with a fraction of the light exiting the cavity at each reflection from the coupling mirror. In our system the spinwave also decays due to decoherence effects, such as control-field induced scattering, which is proportional to the control field power \cite{Hosseini2011}, and atomic diffusion \cite{Zhao2008}.  Similar ring-down dynamics have been observed previously in a Raman atomic memory \cite{Reim2012}.

Fig. \ref{figure2}D shows light accumulation for several control field powers and in Fig. \ref{figure2}E we see the equilibria reached for various control field powers. The best accumulation occurs at low control field power where the coupling in and out of the memory is weaker and the scattering due to the control field is minimised.  As described earlier, this weak coupling condition is analogous to having a ring cavity of the type illustrated in Fig. \ref{figure1}D, with a lower reflectivity coupling mirror giving higher finesse.

\begin{figure}
\includegraphics[width=\textwidth]{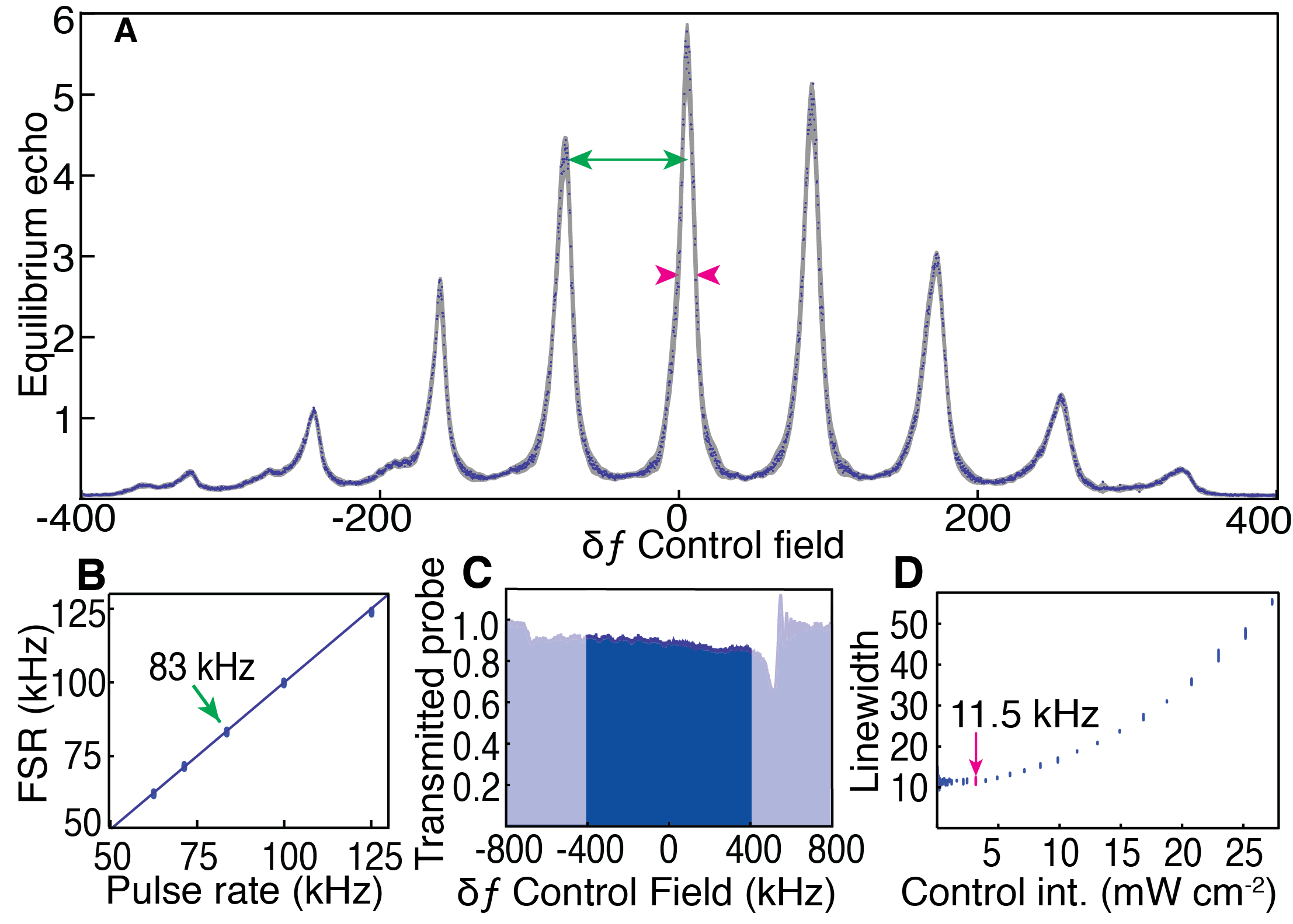}
\caption{\textbf{Resonator behavior of the system. (A)} Amplitude of the equilibrium echo as the control field frequency is scanned 800 kHz over 200 s, sampled every 0.12 kHz. Resonances occur when the phase matching condition is satisfied. Arrows indicate the linewidth and FSR of this spectrum. The shaded area marks the 0.99 confidence interval. \textbf{(B)} Free spectral range for different memory repetition rates.
\textbf{(C)} Broadened Raman absorption through the vapor cell with one gradient coil switched on. The dip and oscillations around 1300 kHz are due to free induction decay. The shaded area indicates the range
over which data were taken. \textbf{(D)} Linewidth of the central resonance for various control field powers. The linewidth reaches a minimum for low control field powers, where diffusion and relaxation losses dominate over coupling and scattering.}\label{figure3}
\end{figure}

The phase of a spinwave depends on the relative phase between the probe and control field at the time of storage. The accumulation of the stored spinwave depends on successive incident pulses contributing to spinwaves of identical phase to give constructive interference. In Fig. \ref{figure3}A, we measure the magnitude of the equilibrium spinwave as we vary the relative phase of successive incident pulses by changing the frequency of the control field. The resonance peaks in Fig. \ref{figure3}A are analogous to those that can be observed in the transmission spectrum of an optical cavity \cite{siegman1986lasers}.

The Free Spectral Range (FSR) of a cavity is determined by the round-trip time for light circulating in the resonator. In our system the effective FSR is determined by the time interval between interference events that couple incident pulses to the spinwave.  In Fig. \ref{figure1}, for example, this interval is 12 $\upmu s$, corresponding to a repetition frequency of 83 kHz.  This is what determines the spacing between the peaks of Fig. \ref{figure3}A.  The condition for resonance is that the phase between successive input spinwaves evolves through an integral number of cycles in the time between coupling events.  By varying the gradient switching period of our experiment, while sending input pulses at a matching period, we find the expected relationship between equivalent Free Spectral Range (FSR) and the interval between gradient switches, as shown in Fig. \ref{figure3}B.

In figure 3A, the peaked Airy function (the square of Eq. \ref{equation1} is spectrally limited, due to the finite bandwidth of the atomic medium that is determined by the magnetic field gradient. This absorption bandwidth is shown in Fig. \ref{figure3}C. A probe pulse is not entirely stored when some of its frequency components lie outside the absorption bandwidth. We use an incident pulse with a bandwidth approximately equal to the bandwidth of the memory. Consequently, only a small region in the center of Fig. \ref{figure3}A is in a regime where all the frequency components of the pulse will be equally absorbed into the memory.

The photon lifetime in an optical resonator determines the width of the Airy peaks \cite{siegman1986lasers}.  Assuming one uses high-quality mirrors to minimize round-trip loss, the only option for decreasing the linewidth is to increase the photon lifetime by building a long cavity, like the 143 m resonator used as mode-cleaner in the VIRGO experiment \cite{TheVirgocollaboration2007}.  The lifetime of photons in a spinwave resonator can be tuned by varying the control field power. Decreasing the control field power lowers the coupling rate of the memory, which is analogous to increasing the cavity finesse. The linewidth of our effective cavity is therefore best in the limit of low control field power, as shown in Fig. \ref{figure3}D.  We note that this is an accurate method to determine the limiting lifetime of our memory, which in this case is found to be $87\pm4\upmu s$ for low control field powers.  The lifetime in this regime has previously been found to be dominated by atomic diffusion \cite{Higginbottom2012}. 

Close to the center of Fig. \ref{figure3}A, we can consider the effective cavity that has been formed in our memory in terms of the characteristics normally measured for optical resonators.  Firstly the FSR is just $83\pm1$ kHz.  This is due to the long storage time of the memory and is equivalent to a cavity with a roundtrip length of about 3.6 km. For small control fields, the linewidth is $11.5\pm0.5$ kHz, giving an equivalent cavity finesse of $7.3\pm0.3$. Considering the frequency of the probe field and the width of the resonance peaks, we find a Q factor of $3.3 (\pm0.1)\times10^{10}$ for storage of the probe field.  In terms of the spinwave, the relevant frequency is the 6.8 GHz ground state splitting of rubidium leading to a spinwave Q factor of $5.9 (\pm0.2)\times10^5$.

\section*{Discussion}
The Q-factor of our system is limited by the diffusion in our atomic ensemble.  This is not an in-principle problem, since both solid state and cold atomic memories have been shown to have much longer coherence times.  Coherence time measurements in cold atomic ensembles \cite{Zhao2008}, silicon-28 \cite{Saeedi2013} and rare-earth systems \cite{Zhong2015} have demonstrated milliseconds, minutes and hours of atomic coherence, respectively. Provided that phase coherence can be maintained, such atomic coherence in combination with the present scheme would lead to effective resonator lengths that are unachievable with standard mirror-based cavities.

Just as a cavity can be used to contain a large resonant optical field, the spinwave resonator can be used to enhance atomic coherences, even in ensembles with low optical depth \cite{Sabooni2013a}.  Using the measured decay rate and assuming the memory absorbs about 30\% of the incident light we can numerically simulate the spinwave (details can be found in the supplementary material).  This shows that repeated constructive interference into the spinwave produces an atomic coherence magnitude about seven times larger than storage of a single pulse.  In fact, this atomic coherence is two times larger than would be achieved with perfect absorption of a single pulse.

The analogy between an optical cavity and our spinwave system is instructive for understanding the behavior of the system, but it is the differences between this setup and a real optical cavity that may give rise to a number of applications. Although we demonstrate a resonator in the classical regime, the linear and noiseless storage properties of the GEM scheme \cite{Hosseini2011a} mean that our system is capable of manipulating and interfering single photon states, making applications in the quantum regime particularly interesting. 

The demonstration of multiple interferences in a memory represents a step towards the realization of a proposal \cite{Campbell2014a}, which relies on using interference between atomic spinwaves and modes of light to implement a memory-based linear-optics quantum processor. In this case the memory would be used to implement multiple interferences between single photon states, as is currently achieved using regular beam splitters \cite{Kok2007}.

Due to the long roundtrip time and pulsed nature of our device, real-time control of the from pulse to pulse is easily achievable. During the 12 µs from one pulse to the next, it is straightforward to measure an echo and conditionally tune both the characteristics of the next pulse and the effective cavity parameters.  For example, by repeatedly sending single photon pulses into the resonator and monitoring the output, Hong-Ou-Mandel interference could be used to conditionally prepare 2-photon Fock states in the spinwave \cite{Xiao2008}.  Higher-order interference between Fock states conditioned on the output of the spinwave, could also be investigated in this way \cite{Sun2007}.

Finally we note that while our realization of a spinwave resonator is based on the GEM protocol, this concept could be extended to other coherent memory techniques. Indeed, ring-down measurements by Reim et al. \cite{Reim2012} indicate that a Raman memory would be suitable.

This research was funded by the Australian Research Council (ARC) Centre of Excellence scheme (CE110001027). BCB and PKL receive financial support from the ARC Future Fellowship scheme.
See Supplement 1 for supporting content.

\bibliographystyle{h-physrev}
\bibliography{Bibliographies-SpinwaveResonator}

\section*{Supplemental Material}
\subsection*{Methods}
The experimental setup is similar to that used in previous work \cite{Hosseini2009,Hosseini2011,Hosseini2011a,Hosseini2012,Sparkes2012a}.  Many of the experimental details can be found in our recent methods paper \cite{Pinel2013}.

The laser source we use is a Ti:Sapphire laser (M-squared Solstis, pumped by Coherent Verdi) around 795 nm, detuned 2 GHz off the $F=2 \rightarrow F’=2$ D1 transition. Part of the light is tapped off to form the control field. Another part is sent through a fiber EOM (EOSPACE), which modulates the light at 6.8 GHz – the ground state hyperfine splitting of $^{87}$Rb.

A ring cavity is used to select the +6.8 GHz sideband of the modulated light. This sideband is used to produce both the probe and local oscillator (LO). Acousto-optic modulators (AOM, model MT80-A1-IR from A.A. Opto-electronic), in a double pass configuration, are used to further tune the frequencies and intensities of the probe, LO and control fields. The LO is detuned from the probe by ~2 MHz, and then recombined with the light coming out of the cell for heterodyne measurement. The probe $e^{-2}$ diameter is 1.25 cm. The LO is larger to minimize noise due to beam pointing.

The probe and control are combined at a 90/10 beamsplitter, keeping most of the control field and discarding most of the probe. A cavity could also be used for lossless combining of the fields. Both fields are carefully circularly polarized using quarter- and half-waveplates and then sent through the cell. The control field is filtered after the rubidium 87 cell using an isotopically enhanced rubidium 85 cell, heated at 150° C – the control field is tuned to be resonant with the $52S1/2 F=3 \rightarrow 52P3/2 F’=2$ transition. The rubidium cell is heated with a non-magnetic resistive wire to a temperature of about $75^\mathrm{o}$C. The heater is switched off during the experiment so the parasitic magnetic fields will not interfere with the experiment.

A constant axial magnetic field is applied to produce a certain level of ground state splitting. Longitudinal magnetic field gradients along the axis are applied using coils with increasing pitch – the frequency of probe at which the Raman transition will occur then depends on the position along the cell. Two coils with opposing gradients are used for storage and recall. The gradient coils are powered with low-voltage power supplies. They are switched using high-speed solid-state switches.

A typical experiment run consisted of sending a series of Gaussian pulses of the probe field through the Rubidium cell. The gradient was switched periodically at twice the pulse repetition rate. 15 pulses were sent, allowing the coherence to accumulate to equilibrium. The equilibrium coherence was measured by taking the final echo, which is proportional to the atomic coherence amplitude at that time. A new experimental run was performed every 60 ms.

\subsection*{Decay and power accumulation measurements} 
The control field frequency was set to maximize the accumulation of atomic coherence inside the cell. Pulse amplitudes were measured by their truncated heterodyne pulse areas, the limit between pulses determined from the minima of the heterodyne signal. The decay data was fitted with exponentials, and the error is the standard deviation of the residuals. The accumulation was normalized to the first echo obtained at each power level. 

\subsection*{Transmission spectra}
The control field was always scanned slowly enough that it could be considered constant over the duration of an experiment run. The amplitude of the last echo was measured from the heterodyne signal by taking the peak-to-peak value of the last echo. Full spectra were obtained by scanning the frequency of the control field by 800 kHz over a period of 200 s. The uncertainty of the measurements was computed from the local RMS noise of the trend-subtracted values.

The spectra of the central peak for linewidth and amplitude measurement were obtained by scanning 100 kHz over 10 s. Each measurement was repeated five times. Airy functions without the assumption of a high finesse cavity were used to fit these spectra. 

The free spectral ranges for various pulse repetition rates were determined from the whole bandwidth data. The interval between peak values was fitted using a least-squares method. FSR error is approximated from phase noise by estimating how far in frequency the measured resonance might be from the actual resonance.

The equations that describe the gradient echo memory used in the experiment can be found in \cite{Hosseini2012}. These equations can be solved numerically. We ran a simulation in one spatial dimension plus time, using XMDS \cite{Dennis2013}. We added an exponential decay term to the atomic coherence to approximate losses due to diffusion.

The results illustrate what occurs inside the memory as pulses are sent in. As the sent pulses are coupled in, the magnitude of the atomic coherence increases. As echo pulses are coupled out, the magnitude decreases. Decay due to other processes is apparent at other times. As the atomic coherence increases, the coupling in also increases due to the interference effect. This is apparent both as the larger increase in the atomic coherence at each coupling and as the decrease in the leakage of the sent pulse. Eventually, the other loss processes balance the increase, and equilibrium is reached.

\begin{figure}
\includegraphics[width=\textwidth]{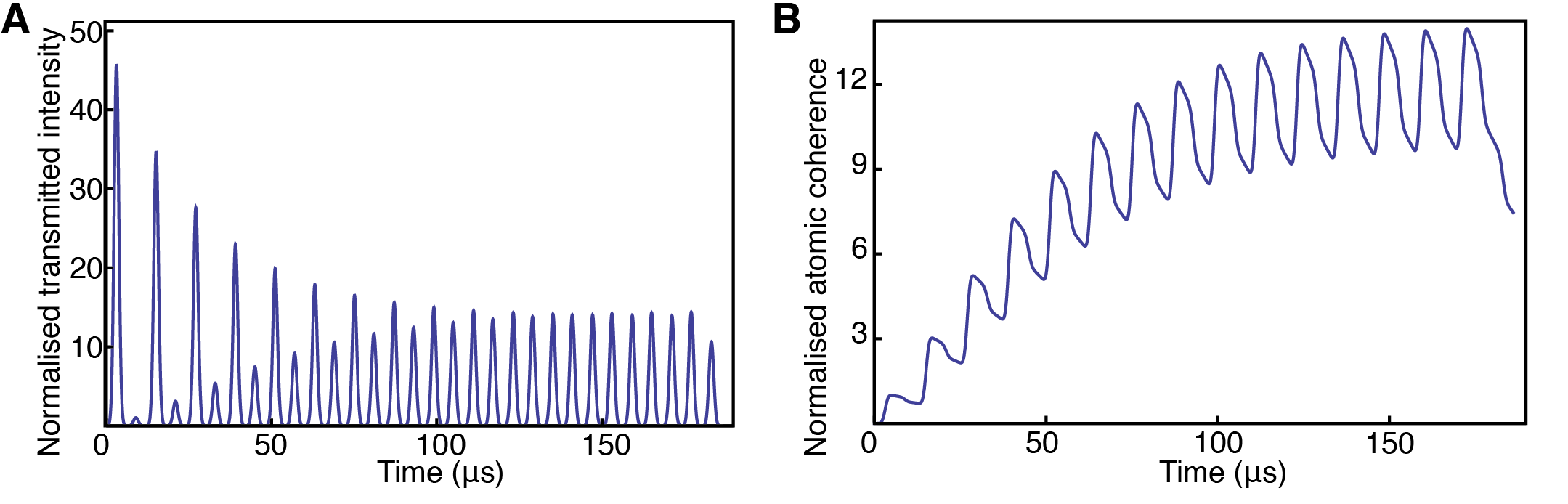}
\caption{Simulation data. \textbf{(A)} Transmitted probe intensity for a series of pulses, scaled against the first echo. Leaked input pulses alternate with recalled pulses. \textbf{(B)} Atomic coherence magnitude for the same simulation. Both graphs are scaled against the first echo/storage in order to show the factor of 14 enhancement in the atomic coherence magnitude.}\label{figure4}
\end{figure}

\end{document}